\documentclass[12pt]{article}
\usepackage[dvips]{color}
\usepackage{epsfig}
\usepackage{amsmath}
\usepackage{graphicx}

\begin{document}
\title{\bf Lagrangian Formulation of a Magnetostatic Field in the Presence of a Minimal Length Scale Based on the Kempf Algebra}

\author{S. K. Moayedi $^{a}$\thanks{E-mail: s-moayedi@araku.ac.ir}\hspace{1mm}
, M. R. Setare $^{b}$ \thanks{E-mail:
rezakord@ipm.ir}\hspace{1mm},
 B. Khosropour $^{a}$\thanks{E-mail: b-khosropour@phd.araku.ac.ir}\hspace{1.5mm}  \\
$^a$ {\small {\em  Department of Physics, Faculty of Sciences,
Arak University, Arak 38156-8-8349, Iran}}\\
$^{b}${\small {\em Department of Science, Campus of Bijar,
University of Kurdistan, Bijar, Iran
}}\\
}
\date{\small{}}
\maketitle
\begin{abstract}
\noindent In the 1990s, Kempf and his collaborators Mangano and
Mann introduced a $D$-dimensional $(\beta,\beta')$-two-parameter
deformed Heisenberg algebra which leads to an isotropic minimal
length $(\triangle
X^{i})_{min}=\hbar\sqrt{D\beta+\beta'}\;,\forall i\in \{1,2,
\cdots ,D\}$. In this work, the Lagrangian formulation of a
magnetostatic field in three spatial dimensions $(D=3)$ described
by Kempf algebra is presented in the special case of
$\beta'=2\beta$ up to the first order over $\beta$. We show that
at the classical level there is a similarity between
magnetostatics in the presence of a minimal length scale
(modified magnetostatics) and the magnetostatic sector of the
Abelian Lee-Wick model in three spatial dimensions. The integral
form of Ampere's law and the energy density of a magnetostatic
field in the modified magnetostatics are obtained. Also, the
Biot-Savart law in the modified magnetostatics is found. By
studying the effect of minimal length corrections to the
gyromagnetic moment of the muon, we conclude that the upper bound
on the isotropic minimal length scale in three spatial dimensions
is $4.42\times10^{-19}m$. The relationship between magnetostatics
with a minimal length and the Gaete-Spallucci non-local
magnetostatics (J. Phys. A: Math. Theor. \textbf{45}, 065401
(2012)) is investigated.

\noindent
\hspace{0.35cm}



\end{abstract}

\section{Introduction}
One of the most important problems in theoretical physics is the
unification between the Einstein's general theory of relativity and
the Standard Model of particle physics [1]. According to Ref. [1],
two important predictions of this unification are the following:
$(i)$ the existence of extra dimensions; and $(ii)$ the existence of
a minimal length scale on the order of the Planck length. Studies in
string theory and loop quantum gravity emphasize that there is a
minimal length scale in nature. Today's theoretical physicists know
that the existence of a minimal length scale leads to a modification
of Heisenberg uncertainty principle. This modified uncertainty
principle can be written as
\begin{equation}
\triangle X \geq\frac{\hbar}{2\triangle
P}+\frac{a_{1}}{2}\;\frac{\ell_{P}^{2}}{\hbar}\;\Delta
P+\frac{a_{2}}{2}\;\frac{\ell_{P}^{4}}{\hbar^{3}}\;(\Delta
P)^{3}+\cdots,
\end{equation}
where $\ell_{P}$ is the Planck length and  $a_{i}\;,\forall i\in
\{1,2, \cdots \}$, are positive numerical constants [2-4]. By
keeping only the first two terms on the right-hand side of Eq.
(1), we obtain the usual generalized uncertainty principle (GUP)
as follows:
\begin{equation}
\triangle X \geq\frac{\hbar}{2\triangle
P}+\;\frac{a_{1}}{2}\;\frac{\ell_{P}^{2}}{\hbar}\;\Delta P.
\end{equation}
It is clear that in Eq. (2), $\triangle X$ is always larger than
$(\triangle X)_{min}=\sqrt{a_{1}}\;\ell_{P}$. At the present time,
theoretical physicists believe that reformulation of quantum field
theory in the presence of a minimal length scale leads to a
divergenceless quantum field theory [5-7]. During recent years,
reformulation of quantum mechanics, gravity, and quantum field
theory in the presence of a minimal length scale have been studied
extensively [5-21]. H. S. Snyder was the first who formulated the
electromagnetic field in quantized spacetime [22]. There are many
papers about electrodynamics in the presence of a minimal length
scale. For a review, we refer the reader to Refs.
[12,13,14,15,16,19,20]. In our previous work [15], we studied
formulation of electrodynamics with an external source in the
presence of a minimal measurable length. In this work, we study
formulation of a magnetostatic field with an external current
density in the presence of a minimal length scale based on the Kempf
algebra. This paper is organized as follows. In Section 2, the
$D$-dimensional $(\beta,\beta')$-two-parameter deformed Heisenberg
algebra introduced by Kempf and his co-workers is studied and it is
shown that the Kempf algebra leads to a minimal length scale
[23-25]. In Section 3, the Lagrangian formulation of a magnetostatic
field in three spatial dimensions described by Kempf algebra is
introduced in the case of $\beta'=2\beta$, whereas the position
operators commute to the first order in $\beta$. It is shown that at
the classical level there is a similarity between magnetostatics in
the presence of a minimal length scale and the magnetostatic sector
of the Abelian Lee-Wick model in three spatial dimensions. The
Ampere's law and the energy density of a magnetostatic field in the
presence of a minimal length scale are obtained. In Section 4, the
Biot-Savart law in the presence of a minimal length scale is found.
We show that at large spatial distances the modified Biot-Savart law
becomes the Biot-Savart law in usual magnetostatics. In Section 5,
we study the effect of minimal length corrections to the
gyromagnetic moment of the muon. From this study we conclude that
the upper bound on the isotropic minimal length scale in three
spatial dimensions is $4.42\times10^{-19}m$. This value for the
isotropic minimal length scale is close to the electroweak length
scale $(\ell_{electroweak}\sim 10^{-18}\, m)$. In Section 6, the
relationship between magnetostatics in the presence of a minimal
length scale and a particular class of non-local magnetostatic field
is investigated. Our conclusions are presented in Section 7. We use
SI units throughout this paper.

\section{Modified Commutation Relations with a Minimal Length Scale}
Kempf and co-workers have introduced a modified Heisenberg algebra
which describes a $D$-dimensional quantized space [23-25]. The Kempf
algebra in a $D$-dimensional space is characterized by the following
modified commutation relations
\begin{eqnarray}
\left[X^{i},P^{j}\right] &=& i\hbar \left[
(1+\beta\textbf{P}^{2})\delta^{ij} +
\beta' P^{i}P^{j}\right], \\
\left[X^{i},X^{j}\right] &=& i\hbar
\frac{(2\beta-\beta')+(2\beta+\beta')\beta \textbf{P}^{2}}{1+\beta
\textbf{P}^{2}}(P^{i}X^{j}-P^{j}X^{i}), \\
\left[P^{i},P^{j}\right] &=& 0,
\end{eqnarray}
where $i,j=1,2,...,D$ and $\beta ,\beta'$ are two non-negative
deformation parameters $(\beta,\beta' \geq 0)$. In Eqs. (3) and
(4), $\beta$ and $\beta'$ are constant parameters with dimension
$(momentum)^{-2}$. Also, in the above equations  $X^{i}$ and
$P^{i}$ are position and momentum operators in the deformed space.\\
An immediate consequence of Eq. (3) is the appearance of an
isotropic minimal length scale which is given by [26]
\begin{equation}
(\triangle X^{i})_{min}=\hbar\sqrt{D\beta+\beta'}\quad ,
\quad\forall i\in \{1,2, \cdots ,D\}.
\end{equation}
In Ref. [27], Stetsko and Tkachuk introduced a representation which
satisfies the modified Heisenberg algebra (3)-(5) up to the first
order in deformation parameters $\beta$ and $\beta'$. The
Stetsko-Tkachuk representations for the position and momentum
operators in the deformed space can be written as follows:
\begin{eqnarray}
X^{i} &=& x^{i}+
\frac{2\beta-\beta'}{4}(\textbf{p}^{2}x^{i}+x^{i}\textbf{p}^{2}), \\
P^{i} &=& p^{i}(1+\frac{\beta'}{2}\textbf{p}^{2}),
\end{eqnarray}
where $x^{i}$ and
$p^{i}={i}{\hbar}\partial^{i}={i}{\hbar}\frac{\partial}{\partial
x_{i}}$ are position and momentum operators in ordinary quantum
mechanics, and $\textbf{p}^{2}=\sum_{i=1}^{D}p^{i}p^{i}$. In this
article, we study the special case of $\beta'=2\beta$, in which
the position operators commute to the first order in deformation
parameter $\beta$, i.e., $[X^{i},X^{j}]=0$ and thus a diagonal
representation for the position operator in the deformed space
can be obtained. For this linear approximation, the modified
Heisenberg algebra (3)-(5) becomes
\begin{eqnarray}
\left[X^{i},P^{j}\right] &=& i\hbar \left[
(1+\beta\textbf{P}^{2})\delta^{ij} +2\beta
 P^{i}P^{j}\right], \\
\left[X^{i},X^{j}\right] &=& 0, \\
\left[P^{i},P^{j}\right] &=& 0.
\end{eqnarray}
In 1999, Brau [28] showed that the following representations satisfy
(9)-(11), in the first order in $\beta$:
\begin{eqnarray}
X^{i} &=& x^{i},\\
P^{i} &=& p^{i}(1+\beta\textbf{p}^{2}).
\end{eqnarray}
It is necessary to note that the Stetsko-Tkachuk representations
(7),(8) and the Brau representations (12),(13) coincide when
$\beta'=2\beta$. Benczik has shown that the energy spectrum of some
quantum systems in the deformed space with a minimal length are
representation-independent [29]. It seems that the laws of physics
in the presence of a minimal length must be
representation-independent.

\section{Lagrangian Formulation of a Magnetostatic Field with an External Current Density in the Presence of a Minimal Length Scale
Based on the Kempf Algebra}

The Lagrangian density for a magnetostatic field with an external
current density $\textbf{J}(\textbf{x})=(J^{1}(\textbf{x}), \\
J^{2}(\textbf{x}), J^{3}(\textbf{x}))$ in three spatial dimensions
$(D=3)$ can be written as follows [30]:
\begin{equation}
{\cal
L}=-\frac{1}{4\mu_{0}}F_{ij}(\textbf{x})F^{ij}(\textbf{x})+J^{i}(\textbf{x})A^{i}(\textbf{x}),
\end{equation}
where $i,j=1,2,3$ ,
$F_{ij}(\textbf{x})=\partial_{i}A_{j}(\textbf{x})-\partial_{j}A_{i}(\textbf{x})$
and
$\textbf{A}(\textbf{x})=(A^{1}(\textbf{x}),A^{2}(\textbf{x}),A^{3}(\textbf{x}))$
are the electromagnetic field tensor and the vector potential
respectively.\\
The Euler-Lagrange equation for the components of the vector
potential is
\begin{equation}
\frac{\partial{\cal L}}{\partial A_{k}
}-\partial_{l}\left(\frac{\partial{\cal
L}}{\partial(\partial_{l}A_{k})}\right)=0.
\end{equation}
If we substitute (14) into (15), we will obtain the following
field equation for the magnetostatic field
\begin{equation}
\partial_{l}F^{lk}(\textbf{x})=\mu_{0}J^{k}(\textbf{x}).
\end{equation}
The electromagnetic field tensor $F_{ij}(\textbf{x})$ satisfies the
Bianchi identity
\begin{equation}
\partial_{i}F_{jk}(\textbf{x})+\partial_{j}F_{ki}(\textbf{x})+\partial_{k}F_{ij}(\textbf{x})=0.
\end{equation}
The three-dimensional magnetic induction vector
$\textbf{B}(\textbf{x})$ is defined as follows [31]:
\begin{equation}
F_{ij}=-\epsilon_{ijk}B^{k}\,\,\,\,
,\,\,\,F^{ij}=\epsilon^{ijk}B_{k}\;,
\end{equation}
where
\begin{equation}
\{B^{i}\}=\{B_{x},B_{y},B_{z}\}\,\,\,\,,\,\,\,\,\{B_{i}\}=\{-B_{x},-B_{y},-B_{z}\}.
\end{equation}
Using Eqs. (18) and (19), Eqs. (16) and (17) can be written in the
vector form as follows:
\begin{eqnarray}
\boldsymbol{\nabla}\times\textbf{B}(\textbf{x})&=&\mu_{0}\textbf{J}(\textbf{x}),\\
\boldsymbol{\nabla}\cdot \textbf{B}(\textbf{x})&=&0.
\end{eqnarray}
The above equations are the basic equations of magnetostatics [30].\\
An immediate consequence of Eq. (21) is that
$\textbf{B}(\textbf{x})$ can be written as follows:
\begin{equation}
\textbf{B}(\textbf{x})=\boldsymbol{\nabla}\times\textbf{A}(\textbf{x}).
\end{equation}
Now, we want to obtain the Lagrangian density for a magnetostatic
field in the peresence of a minimal length scale based on the
Kempf algebra. For this purpose, we must replace the ordinary
position and derivative operators with the deformed position and
derivative operators according to Eqs. (12) and (13), i.e.,
\begin{eqnarray}
x^{i}\longrightarrow  X^{i}&=&x^{i}, \\
\partial^{i}\longrightarrow D^{i}&:=&(1-\beta\hbar^{2}\nabla^{2})\partial^{i},
\end{eqnarray}
where $\nabla^{2}:=\partial_{i}\partial_{i}$ is the Laplace
operator. Using Eqs. (23) and (24) the electromagnetic field tensor
in the presence of a minimal length scale becomes\\
\begin{equation*}
F_{ij}(\textbf{x})=\partial_{i}A_{j}(\textbf{x})-\partial_{j}A_{i}(\textbf{x})\longrightarrow
{\cal F}_{ij}(\textbf{X})=
D_{i}A_{j}(\textbf{X})-D_{j}A_{i}(\textbf{X}),
\end{equation*}
or
\begin{equation}
{\cal
F}_{ij}(\textbf{X})=F_{ij}(\textbf{x})-\beta\hbar^{2}\nabla^{2}F_{ij}(\textbf{x}).
\end{equation}
It should be mentioned that the above modification of the
electromagnetic field tensor has been introduced earlier by
Hossenfelder and co-workers in order to study the minimal length
effects in quantum electrodynamics in Ref. [16]. If we use Eqs.
(23), (24), and (25), we obtain the Lagrangian density for a
magnetostatic field in the deformed space as follows \footnote
{Using Eq. (23) together with the transformation rule for a
contravariant vector, we obtain the following result to the first
order in deformation parameter $\beta$
\begin{eqnarray*}
J^{\;\prime \;i}(\textbf{X})A^{\;\prime
\;i}(\textbf{X})=\frac{\partial X^{i}}{\partial
x^{j}}J^{j}(\textbf{x})\frac{\partial X^{i}}{\partial
x^{k}}A^{k}(\textbf{x})=\delta_{j}^{i}\delta_{k}^{i}J^{
j}(\textbf{x})A^{k}(\textbf{x})=J^{
i}(\textbf{x})A^{i}(\textbf{x}).
\end{eqnarray*}}
:
\begin{eqnarray} {\cal L} &=& -\;\frac{1}{4\mu_{0}}{\cal
F}_{ij}(\textbf{X}){\cal
F}^{ij}(\textbf{X})+J^{i}(\textbf{X})A^{i}(\textbf{X}) \nonumber
\\ &=&
 -\;\frac{1}{4\mu_{0}}
 F_{ij}(\textbf{x})F^{ij}(\textbf{x})+
\frac{1}{4\mu_{0}}(\hbar\sqrt{2\beta})^{2}\nonumber \\
 & &
 F_{ij}(\textbf{x})\nabla^{2}
F^{ij}(\textbf{x})+J^{i}(\textbf{x})A^{i}(\textbf{x})+{\cal
O}\left((\hbar\sqrt{2\beta})^{4}\right).
\end{eqnarray}
The term
$\frac{1}{4\mu_{0}}(\hbar\sqrt{2\beta})^{2}F_{ij}(\textbf{x})\nabla^{2}
F^{ij}(\textbf{x})$ in Eq. (26) can be considered as a minimal
length effect. After neglecting terms of order
$(\hbar\sqrt{2\beta})^{4}$ and higher in Eq. (26) we obtain
\begin{eqnarray}
{\cal L} &=&
-\;\frac{1}{4\mu_{0}}F_{ij}(\textbf{x})F^{ij}(\textbf{x})+
\frac{1}{4\mu_{0}}(\hbar\sqrt{2\beta})^{2}
 F_{ij}(\textbf{x})\nabla^{2}
F^{ij}(\textbf{x})+J^{i}(\textbf{x})A^{i}(\textbf{x}).
\end{eqnarray}
The Lagrangian density (27) is similar to the magnetostatic sector
of the Abelian Lee-Wick model which was introduced by Lee and Wick
as a finite theory of quantum electrodynamics [32-36]. Eq. (27) can
be written as
\begin{eqnarray}
{\cal L} &=&
-\;\frac{1}{4\mu_{0}}F_{ij}(\textbf{x})F^{ij}(\textbf{x})-
\frac{1}{4\mu_{0}}(\hbar\sqrt{2\beta})^{2}
 \partial_{n}F_{ij}(\textbf{x})\,\partial_{n}F^{ij}(\textbf{x})+J^{i}(\textbf{x})A^{i}(\textbf{x}) +
 \partial_{n}\Lambda_{n}(\textbf{x}),
\end{eqnarray}
where
\begin{eqnarray}
\Lambda_{n}(\textbf{x}):=\frac{1}{4\mu_{0}}(\hbar\sqrt{2\beta})^{2}F_{ij}(\textbf{x})
\partial_{n}F^{ij}(\textbf{x}).
\end{eqnarray}
After dropping the total derivative term $
\partial_{n}\Lambda_{n}(\textbf{x})$, the Lagrangian density (28)
will be equivalent to the following Lagrangian density:
\begin{eqnarray}
{\cal L} &=&
-\;\frac{1}{4\mu_{0}}F_{ij}(\textbf{x})F^{ij}(\textbf{x})-
\frac{1}{4\mu_{0}} a ^{2}
\partial_{n}F_{ij}(\textbf{x})\partial_{n}F^{ij}(\textbf{x})+J^{i}(\textbf{x})A^{i}(\textbf{x}),
\end{eqnarray}
where $a:=\hbar\sqrt{2\beta}$ is a constant parameter which is
called Podolsky's characteristic length [37-41]. The
Euler-Lagrange equation for the Lagrangian density (30) is [42-44]
\begin{equation}
\frac{\partial{\cal L}}{\partial A_{k}
}-\partial_{l}\left(\frac{\partial{\cal
L}}{\partial(\partial_{l}A_{k})}\right)+\partial_{m}\partial_{l}\left(\frac{\partial{\cal
L}}{\partial(\partial_{m}\partial_{l}A_{k})}\right)=0.
\end{equation}
If we substitute (30) into (31), we obtain the following field
equation for the magnetostatic field in the deformed space
\footnote{Here, we use the following definition:
\begin{eqnarray*}
\frac{\partial\phi_{i_{1}\cdots i_{k}}}{\partial\phi_{j_{1}\cdots
j_{k}}}=\delta_{i_{1}}^{j_{1}}\cdots\delta_{i_{k}}^{j_{k}},
\end{eqnarray*}
where $\phi_{i_{1}\cdots i_{k}}:=\partial_{i_{1}}\cdots
\partial_{i_{k}}\phi$. This definition has been used by Moeller and Zwiebach in Ref. [44].}

\begin{equation}
\partial_{l}F^{lk}(\textbf{x})-a^{2}\;\nabla^{2}\partial_{l}F^{lk}(\textbf{x})=\mu_{0}J^{k}(\textbf{x}).
\end{equation}
Using Eqs. (18) and (19), Eqs. (17) and (32) can be written in the
vector form as follows:
\begin{eqnarray}
(1-a^{2}\nabla^{2})\boldsymbol{\nabla}\times\textbf{B}(\textbf{x})&=&\mu_{0}\textbf{J}(\textbf{x}),\\
\boldsymbol{\nabla}\cdot\textbf{B}(\textbf{x})&=& 0.
\end{eqnarray}
Equations (33) and (34) are fundamental equations of Podolsky's
magnetostatics [45-48]. It should be noted that Eqs. (30), (33), and
(34) can be obtained as the magnetostatic limit of Eqs. (20), (26),
and (27) in our previous paper [15]. Using Stokes's theorem the
integral form of Eq. (33) can be written in the form:
\begin{equation}
\oint_{C}[\textbf{B}(\textbf{x})-(\hbar\sqrt{2\beta})^{2}\nabla^{2}\textbf{B}(\textbf{x})]\cdot
d\textbf{l}=\mu_{0} I,
\end{equation}
where $I$ is the total current passing though the closed curve $C$.
Equation (35) is Ampere's law in the presence of a minimal length
scale. It is clear that for $\hbar\sqrt{2\beta}\rightarrow 0$, the modified Ampere's law in Eq. (35) becomes the usual Ampere's law. \\
Now, let us obtain the energy density of a magnetostatic field in
the presence of a minimal length scale. The energy density of a
magnetostatic field in the usual magnetostatics is given by [30]
\begin{eqnarray}
u_{B}&=&\frac{1}{2\mu_{0}}\textbf{B}(\textbf{x})\cdot\textbf{B}(\textbf{x})
\nonumber\\
&=&\frac{1}{2\mu_{0}}(\boldsymbol{\nabla}\times\textbf{A}(\textbf{x}))\cdot(\boldsymbol{\nabla}\times\textbf{A}(\textbf{x}))\;.
\end{eqnarray}
Using Eqs. (23) and (24) the energy density of a magnetostatic
field
under the influence of a minimal length scale becomes\\
\begin{eqnarray*}
u_{B}=\frac{1}{2\mu_{0}}(\boldsymbol{\nabla}\times\textbf{A}(\textbf{x}))\cdot(\boldsymbol{\nabla}\times\textbf{A}(\textbf{x}))
\longrightarrow
u^{^{\textbf{ML}}}_{B}=\frac{1}{2\mu_{0}}(\textbf{D}\times\textbf{A}(\textbf{X}))\cdot(\textbf{D}\times\textbf{A}(\textbf{X})),
\end{eqnarray*}
or\\

\begin{eqnarray}
u^{^{\textbf{ML}}}_{B}&=&\frac{1}{2\mu_{0}}[(1-\beta\hbar^{2}\nabla^{2})\boldsymbol{\nabla}\times\textbf{A}(\textbf{x})]\cdot[(1-\beta\hbar^{2}\nabla^{2})
 \boldsymbol{\nabla}\times\textbf{A}(\textbf{x})]
 \nonumber\\
 &=&\frac{1}{2\mu_{0}}\textbf{B}(\textbf{x})\cdot\textbf{B}(\textbf{x})-\frac{1}{2\mu_{0}}(\hbar\sqrt{2\beta})^{2}
 \textbf{B}(\textbf{x})\cdot\nabla^{2}\textbf{B}(\textbf{x})+{\cal
O}\left((\hbar\sqrt{2\beta})^{4}\right),
\end{eqnarray}
where we use the abbreviation $\textbf{ML}$ for the minimal
length.
If we use the vector identities\\
\begin{eqnarray}
\boldsymbol{\nabla}\times(\boldsymbol{\nabla}\times\textbf{a})&=&\boldsymbol{\nabla}(\boldsymbol{\nabla}\cdot\textbf{a})-\nabla^{2}\textbf{a}\;,\\
\boldsymbol{\nabla}\cdot(\textbf{a}\times\textbf{b})&=&\textbf{b}\cdot(\boldsymbol{\nabla}\times\textbf{a})-\textbf{a}\cdot(\boldsymbol{\nabla}\times\textbf{b})\;,
\end{eqnarray}
together with Eq. (34), the modified energy density
$u^{^{\textbf{ML}}}_{B}$ can be written in the form
\begin{eqnarray}
u^{^{\textbf{ML}}}_{B}&=&\frac{1}{2\mu_{0}}\textbf{B}(\textbf{x})\cdot\textbf{B}(\textbf{x})+\frac{1}{2\mu_{0}}(\hbar\sqrt{2\beta})^{2}
(\boldsymbol{\nabla}\times\textbf{B}(\textbf{x}))\cdot(\boldsymbol{\nabla}\times\textbf{B}(\textbf{x}))
\nonumber\\
&&+\boldsymbol{\nabla}\cdot\boldsymbol{\Omega}(\textbf{x}) +{\cal
O}\left((\hbar\sqrt{2\beta})^{4}\right),
\end{eqnarray}
where
\begin{eqnarray}
\boldsymbol{\Omega}(\textbf{x})&:=&\frac{1}{2\mu_{0}}(\hbar\sqrt{2\beta})^{2}
(\boldsymbol{\nabla}\times\textbf{B}(\textbf{x}))\times\textbf{B}(\textbf{x})\;.
\end{eqnarray}
After dropping the total divergence term
$\boldsymbol{\nabla}\cdot\boldsymbol{\Omega}(\textbf{x})$, the
modified energy density (40) will be equivalent to the following
modified energy density:
\begin{eqnarray}
u^{^{\textbf{ML}}}_{B}&=&\frac{1}{2\mu_{0}}\textbf{B}(\textbf{x})\cdot\textbf{B}(\textbf{x})+\frac{1}{2\mu_{0}}(\hbar\sqrt{2\beta})^{2}
(\boldsymbol{\nabla}\times\textbf{B}(\textbf{x}))\cdot(\boldsymbol{\nabla}\times\textbf{B}(\textbf{x}))
\nonumber\\
&&
 +{\cal O}\left((\hbar\sqrt{2\beta})^{4}\right).
\end{eqnarray}
The term $\frac{1}{2\mu_{0}}(\hbar\sqrt{2\beta})^{2}
(\boldsymbol{\nabla}\times\textbf{B}(\textbf{x}))\cdot(\boldsymbol{\nabla}\times\textbf{B}(\textbf{x}))$
in Eq. (42) shows the effect of minimal length corrections.

\section{Green's Function for a Magnetostatic Field in the Presence of a Minimal Length Scale}
Substituting Eq. (22) into Eq. (33) and using the vector identity
(38) we obtain
\begin{eqnarray}
(1-a^{2}\nabla^{2})[\boldsymbol{\nabla}(\boldsymbol{\nabla}\cdot\textbf{A}(\textbf{x}))-\nabla^{2}\textbf{A}(\textbf{x})]=\mu_{0}\textbf{J}(\textbf{x}).
\end{eqnarray}
In the Coulomb gauge
$(\boldsymbol{\nabla}\cdot\textbf{A}(\textbf{x})=0)$, Eq. (43)
can be written as
\begin{eqnarray}
(1-a^{2}\nabla^{2})\nabla^{2}\textbf{A}(\textbf{x})=-\mu_{0}\textbf{J}(\textbf{x}).
\end{eqnarray}
The solution of Eq. (44) in terms of the Green's function,
$G(\textbf{x},\textbf{x}')$, is given by
\begin{eqnarray}
\textbf{A}(\textbf{x})=\textbf{A}_{0}(\textbf{x})+\frac{\mu_{0}}{4\pi}\int
G(\textbf{x},\textbf{x}')\textbf{J}(\textbf{x}')d^{3}x'\; ,
\end{eqnarray}
where $\textbf{A}_{0}(\textbf{x})$ and $G(\textbf{x},\textbf{x}')$
satisfy the equations
\begin{eqnarray}
(1-a^{2}\nabla^{2})\nabla^{2}\textbf{A}_{0}(\textbf{x})=0,
\end{eqnarray}
and
\begin{eqnarray}
(1-a^{2}\nabla^{2}_{\textbf{x}})\nabla^{2}_{\textbf{x}}G(\textbf{x},\textbf{x}')=-4\pi\delta(\textbf{x}-\textbf{x}').
\end{eqnarray}
Now, let us solve Eq. (47) by writting $G(\textbf{x},\textbf{x}')$
and $\delta(\textbf{x}-\textbf{x}')$ in terms of Fourier integrals
as follows:
\begin{eqnarray}
G(\textbf{x},\textbf{x}')&=&\frac{1}{(2\pi)^{3}}\int\;e^{-i\textbf{k}\cdot(\textbf{x}-\textbf{x}')}\widetilde{G}(\textbf{k})d^{3}k\;
,
\\
\delta(\textbf{x}-\textbf{x}')&=&\frac{1}{(2\pi)^{3}}\int\;e^{-i\textbf{k}\cdot(\textbf{x}-\textbf{x}')}d^{3}k\;
.
\end{eqnarray}
If we substitute Eqs. (48) and (49) into Eq. (47), we obtain the
functional form of $\widetilde{G}(\textbf{k})$ as follows:
\begin{eqnarray}
\widetilde{G}(\textbf{k})&=&\frac{4\pi}{\textbf{k}^{2}+a^{2}(\textbf{k}^{2})^{2}}
\nonumber\\
&=&4\pi(\frac{1}{\textbf{k}^{2}}-\frac{a^{2}}{1+a^{2}\textbf{k}^{2}}).
\end{eqnarray}
If Eq. (50) is inserted into Eq. (48), the Green's function
$G(\textbf{x},\textbf{x}')$ becomes
\begin{eqnarray}
G(\textbf{x},\textbf{x}')&=&\frac{1}{2\pi^{2}}\int\;e^{-i\textbf{k}\cdot(\textbf{x}-\textbf{x}')}(\frac{1}{\textbf{k}^{2}}-\frac{a^{2}}{1+a^{2}\textbf{k}^{2}})d^{3}k\;
\nonumber\\
&=&\frac{1-e^{-\frac{|\textbf{x}-\textbf{x}'|}{a}}}{|\textbf{x}-\textbf{x}'|}.
\end{eqnarray}
This type of Green's function has been considered in electrodynamics
to avoid divergences associated with point charges [38,45,49,50].
Using Eqs. (45) and (51) the particular solution of Eq. (44), which
vanishes at infinity is
\begin{eqnarray}
\textbf{A}(\textbf{x})=\frac{\mu_{0}}{4\pi}\int\frac{1-e^{-\frac{|\textbf{x}-\textbf{x}'|}{a}}}{|\textbf{x}-\textbf{x}'|}\textbf{J}(\textbf{x}')d^{3}x'.
\end{eqnarray}
The vector potential (52) satisfies the Coulomb gauge condition
$\boldsymbol{\nabla}\cdot\textbf{A}(\textbf{x})=0$. The
expression (52) can be applied to current circuits by making the
substitution: $\textbf{J}(\textbf{x}')d^{3}x'\rightarrow
Id\textbf{l}'$. Thus
\begin{eqnarray}
\textbf{A}(\textbf{x})=\frac{\mu_{0}
I}{4\pi}\int_{C}\frac{1-e^{-\frac{|\textbf{x}-\textbf{x}'|}{a}}}{|\textbf{x}-\textbf{x}'|}d\textbf{l}',
\end{eqnarray}
where $C$ is the contour defined by the wire. If we use Eqs. (22)
and (52), we obtain the magnetic induction vector
$\textbf{B}(\textbf{x})$ as follows:
\begin{eqnarray*}
\textbf{B}(\textbf{x})=\frac{\mu_{0}}{4\pi}\int\frac{\textbf{J}(\textbf{x}')\times(\textbf{x}-\textbf{x}')}{|\textbf{x}-\textbf{x}'|^{3}}
[1-(1+\frac{|\textbf{x}-\textbf{x}'|}{a})e^{-\frac{|\textbf{x}-\textbf{x}'|}{a}}]d^{3}x',
\end{eqnarray*}
or
\begin{eqnarray}
\textbf{B}(\textbf{x})=\frac{\mu_{0}I}{4\pi}\int_{C}\frac{d
\textbf{l}'\times(\textbf{x}-\textbf{x}')}{|\textbf{x}-\textbf{x}'|^{3}}
[1-(1+\frac{|\textbf{x}-\textbf{x}'|}{a})e^{-\frac{|\textbf{x}-\textbf{x}'|}{a}}].
\end{eqnarray}
Equation (54) is the Biot-Savart law in the presence of a minimal
length scale.\\
In the limit $a=\hbar\sqrt{2\beta}\rightarrow 0$, the modified
Biot-Savart law in (54) smoothly becomes the usual Biot-Savart law,
i.e.,
\begin{eqnarray}
\lim_{a\rightarrow0}\textbf{B}(\textbf{x})=\frac{\mu_{0}I}{4\pi}\int_{C}\frac{d
\textbf{l}'\times(\textbf{x}-\textbf{x}')}{|\textbf{x}-\textbf{x}'|^{3}}.
\end{eqnarray}

\section{Upper Bound Estimation of the Minimal Length Scale in Modified Magnetostatics}
Now, let us estimate the upper bounds on the isotropic minimal
length scale in modified magnetostatics. By putting $\beta'=2\beta$
into (6) the isotropic minimal length scale becomes
\begin{equation}
(\triangle X^{i})_{min}=\sqrt{\frac{D+2}{2}}\;(\hbar\sqrt{2\beta})\;
, \quad\forall i\in \{1,2, \cdots ,D\}.
\end{equation}
The isotropic minimal length scale (56) in three spatial dimensions
is given by
\begin{equation}
(\triangle X^{i})_{min}={\frac{\sqrt{10}}{2}}\;a\quad , \quad\forall
i\in \{1,2,3\},
\end{equation}
where $a=\hbar\sqrt{2\beta}$.\\
In a series of papers, Sprenger and co-workers [51,52] have
concluded that the minimal length scale $(\triangle X^{i})_{min}$ in
Eq. (57) might lie anywhere between the Planck length scale
$(\ell_{P}\sim 10^{-35}\, m)$ and the electroweak  length scale
$(\ell_{electroweak}\sim 10^{-18}\, m)$, i.e.,
\begin{equation}
 10^{-35}\, m < (\triangle X^{i})_{min}< 10^{-18}\, m.
\end{equation}
According to above statements, the upper bound on the isotropic
minimal length scale in three spatial dimensions becomes
\begin{equation}
 (\triangle X^{i})_{min}< 10^{-18}\, m.
\end{equation}
Inserting (59) into (57), we find
\begin{equation}
 a < 0.63 \times 10^{-18}\, m.
\end{equation}
In a series of papers, Accioly et al. [34, 36, 37] have estimated an
upper bound on Podolsky's characteristic length $a$ by computing the
anomalous magnetic moment of the electron in the framework of
Podolsky's electrodynamics. This upper bound on $a$ is
\begin{equation}
 a < 4.7 \times 10^{-18}\, m.
\end{equation}
Note that the upper bound on the Podolsky's characteristic length
$a$ in Eq. (60) is near to the upper bound on the Podolsky's
characteristic length in Eq. (61).\\
Another upper bound on the minimal length scale has been obtained in
Ref. [53] by considering minimal length corrections to the
gyromagnetic moment of electrons and muons. If we compare Eq. (13)
in this work with Eq. (40) in Ref. [16], we obtain
\begin{equation}
\hbar\sqrt{\beta}=\frac{L_{f}}{\sqrt{3}}\;,
\end{equation}
where $L_{f}$ is the minimal length scale in Refs. [16,53]. If we
substitute (62) into (56), we will obtain the isotropic minimal
length in three spatial dimensions as follows:
\begin{equation}
 (\triangle X^{i})_{min}=\sqrt{\frac{5}{3}}L_{f},\quad\forall
i\in \{1,2,3\}.
\end{equation}
The minimal length scale $L_{f}$ in Eqs. (62) and (63) can be
written as
\begin{equation}
L_{f}=\frac{\hbar}{M_{f}c}\;,
\end{equation}
where $M_{f}$ is a new fundamental mass scale [16,53]. Inserting Eq.
(64) into Eq. (63), we find
\begin{equation}
 (\triangle X^{i})_{min}=\sqrt{\frac{5}{3}}\; \frac{\hbar}{M_{f}c}\;,\quad\forall
i\in \{1,2,3\}.
\end{equation}
In Ref. [53] it was shown that the effect of minimal length
corrections to the gyromagnetic moment of the muon leads to the
following lower bound on the fundamental mass scale of the theory:
\begin{equation}
 M_{f}\geq 577\; \frac{GeV}{c^{2}}.
\end{equation}
Substituting Eq. (66) into Eq. (65), the isotropic minimal length
scale in three spatial dimensions becomes
\begin{equation}
 (\triangle X^{i})_{min}\leq 4.42\times10^{-19}m.
\end{equation}
If we insert Eq. (67) into Eq. (57), we will find
\begin{equation}
 a \leq 2.79\times 10^{-19}\, m.
\end{equation}
It is interesting to note that the numerical value of the upper
bound on $a$ in Eq. (68) and the numerical value of the upper
bound on $a$ in Eq. (60) are close to each other.

\section{The Equivalence between the Gaete-Spallucci Non-Local Magnetostatics and Magnetostatics in the Presence of a Minimal Length Scale}
Smailagic and Spallucci have proposed an approach to formulate
quantum field theory in the presence of a minimal length scale
[54-56]. Using the Smailagic-Spallucci approach, Gaete and Spallucci
have introduced a $U(1)$ gauge field with a non-local kinetic term
whose magnetostatic sector is
\begin{equation}
{\cal L}=-\frac{1}{4\mu_{0}}F_{ij}(\textbf{x})exp(-\theta
\nabla^{2})F^{ij}(\textbf{x})+J^{i}(\textbf{x})A^{i}(\textbf{x}),
\end{equation}
where $\theta$ is a constant parameter with dimension of
$(length)^{2}$ [57]. The function $exp(-\theta\nabla^{2})$ in Eq.
(69) can be expanded in a formal power series as follows:
\begin{equation}
\exp{(-\theta\nabla^{2})}=\sum_{l=0}^{+\infty}
(-1)^{l}\;\frac{\theta^l}{l\;!}\;(\nabla^{2})^{l},
\end{equation}
where $(\nabla^{2})^{l}$ denotes the $\nabla^{2}$ operator applied
$l$ times [58].\\
After inserting Eq. (70) into Eq. (69), we obtain the following
Lagrangian density:
\begin{eqnarray}
{\cal
L}&=&-\frac{1}{4\mu_{0}}F_{ij}(\textbf{x})F^{ij}(\textbf{x})+\frac{1}{4\mu_{0}}\theta
F_{ij}(\textbf{x})\nabla^{2}F^{ij}(\textbf{x})\nonumber\\
 &+&\frac{1}{4\mu_{0}}\sum_{l=2}^{+\infty}
(-1)^{l+1}\;\frac{\theta^l}{l\;!}\;F_{ij}(\textbf{x})(\nabla^{2})^{l}F^{ij}(\textbf{x})+J^{i}(\textbf{x})A^{i}(\textbf{x}).
\end{eqnarray}
If we neglect terms of order $\theta^{2}$ and higher in Eq. (71), we
find
\begin{equation}
{\cal
L}=-\frac{1}{4\mu_{0}}F_{ij}(\textbf{x})F^{ij}(\textbf{x})+\frac{1}{4\mu_{0}}\theta
F_{ij}(\textbf{x})\nabla^{2}F^{ij}(\textbf{x})+J^{i}(\textbf{x})A^{i}(\textbf{x}).
\end{equation}
A comparison between Eqs. (27) and (72) clearly shows that there
is an equivalence between the Gaete-Spallucci non-local
magnetostatics to the first order in $\theta$ and the
magnetostatic sector of the Abelian Lee-Wick model (or
magnetostatics in the presence of a minimal length scale). The
relationship between the non-commutative constant parameter
$\theta$ in Eq. (72) and $a=\hbar\sqrt{2\beta}$ in Eq. (27) is
\begin{equation}
\theta=a^{2}.
\end{equation}
According to Eq. (73), $a=\sqrt{\theta}$ plays the role of the
minimal length in the Gaete-Spallucci non-local magnetostatics
[57,59].\\
If we insert Eq. (73) into Eq. (57), we find
\begin{equation}
(\triangle X^{i})_{min}=\frac{\sqrt{10\;\theta}}{2}\quad ,
\quad\forall i\in \{1,2,3\}.
\end{equation}
Using Eq. (68) in Eq. (73), we obtain the following upper bound
for the non-commutative parameter $\theta$:
\begin{equation}
\theta_{_{\textbf{MLCGMM}}}\leq 7.78\times10^{-38} \; m^{2},
\end{equation}
where we use the abbreviation $\textbf{MLCGMM}$ for the minimal
length corrections to the gyromagnetic moment of the muon. Chaichian
and his collaborators have investigated the Lamb shift in
non-commutative quantum electrodynamics $(\textbf{NCQED})$ [60,61].
They found the following upper bound for the non-commutative
parameter $\theta$:
\begin{eqnarray*}
\theta_{_{\textbf{NCQED}}}\leq(10^{4} GeV)^{-2} ,
\end{eqnarray*}
or
\begin{equation}
\theta_{_{\textbf{NCQED}}}\leq 3.88\times10^{-40}\;m^{2} .
\end{equation}
For a review of the phenomenology of non-commutative geometry see
Ref. [62]. The upper bound (75) is about two orders of magnitude
larger than the upper bound (76), i.e.,
\begin{equation}
\theta_{_{\textbf{MLCGMM}}}\sim \;10^{2}
\;\theta_{_{\textbf{NCQED}}}.
\end{equation}
If we insert (61) into (73), we obtain the following upper bound
for $\theta$:
\begin{equation}
\theta_{_{\textbf{MLCGME}}}\leq 2.2\times10^{-35} \; m^{2},
\end{equation}
where we use the abbreviation $\textbf{MLCGME}$ for the minimal
length corrections to the gyromagnetic moment of the electron.
The upper bound (78) is about four orders of magnitude larger than
the upper bound (76), i.e.,
\begin{equation}
\theta_{_{\textbf{MLCGME}}}\sim \;10^{4}
\;\theta_{_{\textbf{NCQED}}}.
\end{equation}
A comparison between Eq. (77) and Eq. (79) shows that
$\theta_{_{\textbf{MLCGMM}}}$ is nearer to
$\theta_{_{\textbf{NCQED}}}$. It should be emphasized that the
magnetostatics in the presence of a minimal length scale is only
correct to the first order in the deformation parameter $\beta$,
while the Gaete-Spallucci non-local magnetostatics is valid to
all orders in the non-commutative parameter $\theta$.

\section{Conclusions}
After the appearance of quantum field theory many theoretical
physicists have attempted to reformulate quantum field theory in the
presence of a minimal length scale [63,64]. The hope was that the
introduction of such a minimal length scale leads to a
divergenceless quantum field theory [65]. Recent studies in
perturbative string theory and quantum gravity suggest that there is
a minimal length scale in nature [1]. Today's we know that the
existence of a minimal length scale leads to a generalization of
Heisenberg uncertainty principle. An immediate consequence of the
GUP is that the usual position and derivative operators must be
replaced by the modified position and derivative operators according
to Eqs. (23) and (24) for $\beta'=2\beta$. We have formulated
magnetostatics in the presence of a minimal length scale based on
the Kempf algebra. It was shown that there is a similarity between
magnetostatics in the presence of a minimal length scale and the
magnetostatic sector of the Abelian Lee-Wick model. The integral
form of Ampere's law and the energy density of a magnetostatic field
in the presence of a minimal length scale have been obtained. Also,
the Biot-Savart law in the presence of a minimal length scale has
been found. We have shown that in the limit
$\hbar\sqrt{2\beta}\rightarrow 0$, the modified Ampere and
Biot-Savart laws become the usual Ampere and Biot-Savart laws. It is
necessary to note that the upper bounds on the isotropic minimal
length scale in Eqs. (59) and (67) are close to the electroweak
length scale $(\ell_{electroweak}\sim 10^{-18}\, m)$. We have
demonstrated the equivalence between the Gaete-Spallucci non-local
magnetostatics up to the first order over $\theta$ and
magnetostatics with a minimal length up to the first order over the
deformation parameter $\beta$. Recently, Romero and collaborators
have formulated a higher-derivative electrodynamics [66]. In this
work we have formulated a higher-derivative magnetostatics in the
framework of Kempf algebra whereas the authors of [66] have studied
an electrodynamics consistent with anisotropic transformations of
spacetime with an arbitrary dynamic exponent $z$.

\vspace{2cm}


\section*{Acknowledgments}
We are grateful to S. Meljanac and J. M. Romero for their interest
in this work and for drawing our attention to the references
[8,19,66]. Also, we would like to thank the referee for useful
comments and suggestions.

\end{document}